\documentclass[twocolumn,showpacs,preprintnumbers,amsmath,amssymb,nofootinbib,superscriptaddress]{revtex4}
\usepackage{hyperref}
\usepackage{graphicx}
\usepackage{color}

\newcommand{\be}{\begin{equation}}  
\newcommand{\ee}{\end{equation}}  
\newcommand{\bear}{\begin{eqnarray}}  
\newcommand{\eear}{\end{eqnarray}}  
\newcommand{\ba}{\begin{array}}  
\newcommand{\ea}{\end{array}}

\newcommand{\z }{}

\newskip\humongous \humongous=0pt plus 1000pt minus 1000pt

\newif\ifdtup

\def\oldreffmt#1{\rlap{[#1]} \hbox to 2\parindent{}}

\def\figfmt#1{\rlap{Figure {#1}} \hbox to 1in{}}  
  
\def\ie{\hbox{\it i.e.}{}}	  
\def\eg{\hbox{\it e.g.}{}}

\def\VEV#1{\left\langle #1\right\rangle}

\def\slash#1{#1\!\!\!/\!\,\,}  
\def\beq{\begin{equation}}  
\def\eeq{\end{equation}}  
\def\bea{\begin{eqnarray}}  
\def\eea{\end{eqnarray}}  
\def\half{\frac{1}{2}}  
  
\def\bq{\begin{quote}}  
\def\eq{\end{quote}}

\def\half{\frac{1}{2}}       
  
\def \gta {\mathrel{\vcenter  
     {\hbox{$>$}\nointerlineskip\hbox{$\sim$}}}}   

\relax  

\newdimen\tdim  
\tdim=\unitlength  
\def\bar{\overline}

\begin{document}  

\preprint{FERMILAB-PUB-14-086-T}
\title{Ultra-weak sector, Higgs boson mass, 
and the dilaton}

\author{Kyle Allison}
\affiliation{Department of Theoretical Physics\\
University of Oxford, 1 Keble Road\\
Oxford OX1 3NP\\$ $}

\author{\\Christopher T. Hill}
\email{hill@fnal.gov}
\affiliation{Fermi National Accelerator Laboratory\\
P.O. Box 500, Batavia, Illinois 60510, USA\\$ $}

\author{\\Graham G. Ross}
\affiliation{Department of Theoretical Physics\\
University of Oxford, 1 Keble Road\\
Oxford OX1 3NP\\$ $}

\date{\today}

\begin{abstract}
The Higgs boson mass may arise from a portal coupling to
a singlet field $\sigma$ which has a very large VEV $f\gg m_\text{Higgs}$.
This requires a sector of ``ultra-weak'' couplings $\zeta_i$, where $\zeta_i \lesssim m_\text{Higgs}^2/ f^2$. Ultra-weak couplings are technically naturally small due to a custodial \z{shift symmetry  of $\sigma$ in the $\zeta_i\rightarrow 0$ limit.} The singlet field $\sigma$ has properties similar to a pseudo-dilaton.
We engineer explicit breaking of scale invariance in the ultra-weak sector
via a Coleman-Weinberg potential, which requires hierarchies amongst the 
ultra-weak couplings. 
\end{abstract}

\maketitle

\noindent
{\bf I. Introduction}
\vskip .1in
\noindent

The Higgs boson presents several well-known puzzles associated
with the problem of the naturalness of the existence
of a low mass fundamental $0^+$ field in quantum field theory.
The naturalness issue is associated with how scale symmetry is implemented (or not) for the Higgs boson, and there has been a recent upsurge of interest in
models that attempt to maintain a classical scale invariance which is broken only
by scale anomalies \cite{bardeen,general,Iso,CTH}. 
\z{Here we explore this idea in the context of an extension of the Standard Model (SM) that includes a new gauge singlet scalar field $\sigma$ coupled to the Higgs sector via ultra-weak couplings. In particular, we  assume that the Higgs  couples to
the singlet field $\sigma$ through a portal interaction $\zeta_1 \sigma^2 H^\dagger H$. Electroweak breaking is induced when $\sigma$  acquires a VEV by 
quantum loops, \ie, through Coleman-Weinberg (CW)
symmetry breaking \cite{CW}, and thus yields a mass for $\sigma$ and for the Higgs boson.
We consider the case that the $\sigma$ field
VEV $f$ is much larger than the weak scale, $f \gg v_{weak}$, in which case the coupling $\zeta_1$  must be ultra-weak, $|\zeta_1| = m_H^2/f^2 \ll 1$.}
 
At first sight, constructing a model with ultra-weak scalar couplings would seem to be a foolish thing to do since most SM couplings are either technically naturally small
(\eg, the electron or up and down quark Higgs-Yukawa couplings) or are of
order the gauge couplings, such as $g_{top} \sim g_3$.  For example,
the Higgs quartic coupling $\lambda$ receives additive contributions from the large $O(1)$
couplings $g_{top}$, $g_2$ and $g_1$, and thus $\lambda$ is not ultra-weak.

Therefore we must ask if  $\zeta_1$ can be technically naturally small.
The answer is yes:  there exists a {\em custodial symmetry}  for ultra-weak couplings
amongst singlet fields. This is a ``shift symmetry'' and it has a Noether current whose
divergence is small, $\propto \zeta_i$.  This is the reason why ultra-weak couplings can
remain ultra-weak in the renormalization group (RG) evolution; 
the 't Hooft naturalness of ultra-weak couplings is the exact shift symmetry in the limit $\zeta_i\rightarrow 0$. We have seen shift symmetry in another guise before. Shift symmetry naturally casts $\sigma$ as a pseudo-dilaton (see Appendix). 

As discussed in a companion paper \cite{AHR}, one motivation for such small couplings arises in the context of the DFSZ axion solution  to the strong CP problem where $f$ is identified with the axion decay constant \cite{DFS}. Alternatively, it may be that $\sigma$ is the dilation responsible for generating the Planck scale, or $f$ may be
associated with a high energy scale such as Grand Unification. In this paper, we wish to demonstrate, in the context of a very simple model, how such small couplings are natural and to briefly explore the new phenomenology associated with ultra-weak couplings.

\vskip .2in
\noindent
{\bf II. Origin of Higgs boson mass from an ultra-weak sector}
\vskip .1in
\noindent

\z{Consider an extension of the SM in which the Higgs boson sector includes a real singlet scalar field $\sigma$. We assume that the theory has a classical scale symmetry\footnote{\z {The idea of classical scale symmetry as
a custodial symmetry of a fundamental perturbative Higgs boson has been
emphasized by Bardeen \cite{bardeen} who argues that the additive quadratic divergences associated, for example, with the top quark loop correction, are an artifact of using a calculational method that violates classical scale invariance. We will return to a discussion of this below.} } so that only dimension four terms are allowed, giving the most general action of the form}
\beq
\label{theory}
S=\int d^{4}x \left( \frac{1}{2}\partial _{\mu }\sigma \partial
^{\mu }\sigma + 
(D_\mu H)^\dagger D^\mu H - V(H, \sigma) \right),
\eeq
where 
\beq
\label{V1}
V(H, \sigma)=  \frac{\lambda}{2}(H^\dagger H)^2+ \frac{\zeta_1}{2}\sigma^2 H^\dagger H + \frac{\zeta_2}{4}\sigma^4.
\eeq
CW symmetry breaking is  analogous to a QCD-like mechanism  in that it  arises entirely from a stress-tensor trace anomaly, \ie, it relies upon scale symmetry breaking by perturbative quantum loops. The scale invariance of the action is again recovered in the limit $\hbar \rightarrow 0$ as quantum loops are turned off and the trace anomaly goes to zero.  

The RG equations for eq.(\ref{V1}) are
\bea
\label{RGs}
\beta_\lambda  & = & \frac{d\lambda (\mu)}{d\ln (\mu)}
= \frac{1}{ 16\pi^2 } \left( 12\lambda^{2}-3\lambda (3g_2^2+g_1^2) \phantom{\frac{1}{1}}\right.
\nonumber \\
& & \!\!\!\!\!\!   \!\!\!\!\!\!   \left.+ \frac{3}{4}(g_1^2 + g_2^2)^2  +\frac{3}{2}g_2^4 + 12\lambda g_t^2  - 12 g_t^4 +\zeta_1^2 \right),
 \\
\beta_1 & = &  \frac{d\zeta_1 (\mu)}{d\ln (\mu)} = \frac{1}{ 16\pi^2 }\left( 6\zeta_1 \zeta_2 
+ 6\zeta_1\lambda + 4\zeta^2_1 \phantom{\frac{1}{1}} \right.
\nonumber \\
& & \left. -\frac{3}{2}\zeta_1 (3g_2^2+g_1^2)+ 6\zeta_1 g_t^2\right),
 \\
\!\!\!\!\!\!  \!\!\!\!\!\!  
\beta_2 & = & \frac{d\zeta_2 (\mu)}{d\ln (\mu)} = \frac{1}{ 16\pi^2 } \left( 18\zeta_2^2 + 2\zeta_1^2 \right). \label{eq:b2}
\eea
An immediately obvious feature is that\z{, due to the custodial shift symmetry of $\sigma$,} the $\zeta_i$ couplings, as a class, are
multiplicatively renormalized.  Therefore if these couplings are very small they will remain small over a large range of RG running, \ie,  an ``ultra-weak sector'' 
can be technically natural in the SM.

Let us neglect the contribution of the Higgs field to the potential momentarily. The $\sigma$ field, despite having ultra-weak couplings, can 
have a nontrivial CW potential with a minimum
at some high energy scale $f$ \cite{CW}. 
This requires that:
(1) the quartic coupling $\zeta_2$ is negative for any scale $\mu \leq f$,
(2) $\beta_2$ is positive, and (3) $\zeta_2(f') = 0$ at some scale $f' \gta f$ so that
$\zeta_2$ crosses from negative to positive values
with increasing $\ln(\mu)$~\cite{CTH}. The solution to eq.(\ref{eq:b2}) is, to a
 good approximation, $\zeta_2(\mu)\approx \beta_2\ln(\mu/f') $
with constant $\beta_2>0$. This solution satisfies conditions (1) through (3), which can be consistent with the overall quartic stability of the potential.

Using the approximate solution for $\zeta_2(\mu)$ with the VEV of $\sigma$ itself as the scale $\mu$, the effective potential $V_{CW}(\sigma)$ for the field $\sigma$ is
\beq
V_{CW}(\sigma)\approx \frac{1}{4} \beta_2\sigma^4 \ln\left(\frac{\sigma}{f'}\right).
\eeq
The minimum occurs at $\VEV{\sigma} \equiv f=f'e^{-1/4}$.
We also see that
\beq
\label{oneten0}
\beta_2(f) =-4\zeta_2(f),
\eeq
which is the key ``extremal relationship'' 
at the minimum of the CW potential \cite{CTH}. Note that the extremal condition eq.(\ref{oneten0}) says that we are equating a one-loop ${\cal{O}}(\hbar)$ expression, $\beta_2$,
to a tree-level (classical) coupling, $\zeta_2$ \cite{CW}. 
The consistency of this result with perturbation theory requires that the $\zeta_1^2$ term in $\beta_2$ be the dominant one, so
\bea
\label{extremal}
\beta_2(f)\approx \frac{\zeta_1^2(f)}{8\pi^2}, \;\;\;
\makebox{hence}  \;\;\;
 \frac{\beta_2(f)}{4|\zeta_2(f)|}  & \approx &  \frac{\zeta_1^2(f)}{32\pi^2 |\zeta_2(f)|} = 1.
\nonumber \\
\eea
Thus the consistency of the CW potential minimum
requires a substantial hierarchy $|\zeta_2| \ll |\zeta_1| \ll 1$ amongst the
ultra-weak couplings.

For our present problem, however, we have a mixed potential involving the Higgs and $\sigma$ fields,
\beq
\label{mixed}
V(\sigma, H)\approx \frac{1}{4} \beta_2\sigma^4 \ln\left(\frac{\sigma}{{f}'}\right)
+ \frac{\lambda}{2}(H^\dagger H)^2+ \frac{\zeta_1}{2}\sigma^2 H^\dagger H.
\eeq
The minimization procedure can be simplified by writing this as a sum
of two independent potentials,
\beq
\label{newpot}
V(\sigma, H)\approx \frac{1}{4} \beta_2\sigma^4\ln\left(\frac{\sigma}{\tilde{f}}\right)
+ \frac{\lambda}{2}\left(H^\dagger H- \epsilon \sigma^2\right)^2,
\eeq
where
\beq
\label{epsdef}
\epsilon = \frac{|\zeta_1|}{2\lambda}, \qquad f'  = \tilde{f}\exp\left(-\frac{\zeta^2_1}{2\lambda\beta_2} \right),
\eeq
and $\zeta_1$ is negative.

Note that we have {\em not chosen a new RG trajectory} parametrized by
$\tilde{f}$.  Instead, the CW potential appearing in eq.(\ref{newpot}) involves $\tilde{f}$ which is related to our original choice
of trajectory (parametrized by $f'$) through the relationship
\beq
 \beta_2\ln({{f}'})=
 \beta_2\ln({\tilde{f}}) - {2\lambda\epsilon^2}.
\eeq
The zero-crossing of the original  $\zeta_2(\mu)=\beta_2\ln(\mu/{{f}'}) $ 
remains at ${f}'$. However, what now matters
for the minimization of eq.(\ref{newpot}) is  the running of an effective shifted coupling
constant $\zeta'_2(\mu) =\beta_2\ln(\mu/{\tilde{f}})  = \zeta_2(\mu)-{2\lambda\epsilon^2}$. 
Using eqs.(\ref{extremal},\ref{epsdef}), $\zeta_2'(\mu)$ will have a
zero-crossing at a much higher energy scale $\tilde{f} = f' \exp(4\pi^2/\lambda)$,
but it can readily satisfy the extremal condition eq.(\ref{oneten0}) at $\tilde{f} $.  

The minimum of eq.(\ref{newpot}) now occurs at
\bea
\VEV{\sigma} & \equiv & f = \tilde{f}e^{-1/4}, \\
(\VEV{H})^2 & \equiv & v^2 = {\epsilon}f^2, \;\;\;\makebox{hence}\;\;\; \epsilon = \frac{v^2}{f^2}.
\eea
The mass eigenstates are computed by expanding the fields about the minimum,
$\sigma = f + \hat{\sigma}$ and $H = v + h/\sqrt{2}$ ($H$ can be
treated like a complex singlet at this point by going to unitary gauge).
The quadratic terms in the potential, $V(\hat\sigma,h)_2$, are then
\begin{align}
&V(\hat\sigma,h)_2 = 
\half \left(\beta_2 f^{2}+4\lambda\frac{v^4}{f^2}\right) \hat\sigma^2  
+ {\lambda}v^2h^2  - \frac{2\sqrt{2}\lambda v^3}{f} h\hat\sigma 
\nonumber \\
&\quad = 
\half \left(\beta_2 +4\lambda\epsilon^2\right)f^{2} \hat\sigma^2  
+ {\lambda}\epsilon f^2h^2  - 2\sqrt{2}\lambda \epsilon^{3/2}f^2 h \hat\sigma.
\end{align}
Denoting the physical mass eigenstates as $\tilde{h}$ and $\tilde\sigma$,
we find
\beq
h = \tilde{h} + \frac{\sqrt{2}v\tilde\sigma}{f},
\qquad
\sigma = \tilde{\sigma} -\frac{\sqrt{2}v\tilde{h}}{ f}.
\eeq 
To leading order in $\epsilon$, the eigenfields are diagonal with masses
\begin{align}
m_h^2 &= 2\lambda v^2 = 2\lambda\epsilon { f^2} \; \leftrightarrow \;\;\tilde{h},\\
m_\sigma^2 &= \beta_2 f^{2} \; \leftrightarrow \;\; \tilde{\sigma}.
\end{align}
Our model is predictive in terms of $\epsilon = v^2/f^2$
given that, from eqs.(\ref{extremal},\ref{epsdef}), we have
\beq
\label{eq:mSigmaSq}
\beta_2 = \frac{\zeta_1^2}{8\pi^2} = \frac{\lambda^2\epsilon ^2}{2\pi^2},
\qquad\makebox{therefore}\;\;\;  m_\sigma^2  =    \frac{m_h^4}{8\pi^2 f^2}.
\eeq
For  $m_h = 126$ GeV, eq.(\ref{eq:mSigmaSq}) gives
\beq
\label{eq:msigma}
m_\sigma \approx 0.179 \left(\frac{10^{10}\text{~GeV}}{f}\right)\;\makebox{keV}.
\eeq
The model therefore predicts a low mass $0^+$ particle for $f\gta 10^{10}$~GeV. 

The field $\tilde\sigma$ is effectively a dilaton and couples to
everything the Higgs does with the replacement of the Higgs VEV
\beq
v\rightarrow v + \frac{\sqrt{2}v}{f}\tilde\sigma, \;\;\; \makebox{hence} \;\;\; \frac{\delta v^2}{v} = 
\frac{2\sqrt{2} }{f}\tilde\sigma.
\eeq
For example, $\tilde\sigma$ couples to the electron as
\beq
{\cal{L}}' = -\frac{\sqrt{2}m_e\tilde\sigma}{f}\bar{\psi\psi}.
\eeq
Furthermore, this implies that $\tilde{\sigma}$ will couple to the electromagnetic field
$F_{\mu\nu}F^{\mu\nu}$ through  vacuum polarization loops
of all the charged particles in the SM.  This coupling is determined
by the QED $\beta$-function and  satisfies the familiar dilaton low energy theorems
that apply to a very low mass Higgs boson \cite{SVVZ,Spira}.

The $\tilde\sigma \rightarrow \gamma\gamma$ decay width
can by determined by rescaling eq.(1) of ref.\cite{Spira}, giving
\bea
 \Gamma(\tilde\sigma \rightarrow \gamma\gamma) 
& = & C^2_\sigma\frac{\alpha^2 m_\sigma^3 }{256\pi^3f^2},
\eea
with the coefficient $C_\sigma$ given by
\beq
C_\sigma = \left(\sum_Q e_Q^2N_c A_f(0) +  A_W(0) \right) = \frac{11}{3},
\eeq 
where $A_f(0)= 4/3$, $A_W(0)=-7$, and the sum over $Q$ extends over {\em all charged 
fermions} in the SM, yielding $\sum_Q e_Q^2N_c = 8$.\footnote{Here the low energy
theorem is almost exact, in contrast to
the Higgs case for which the sum includes only the top quark and $W$ loops
and the functions $A_f(\tau_f)$ and $A_W(\tau_W)$ in eq.(2) of 
ref.\cite{Spira} are evaluated at nonzero $\tau_i \propto m_h^2/v^2$.}
Using eq.(\ref{eq:msigma}), this leads to a lifetime for the mass eigenstate $\tilde \sigma$ of
\bea
\nonumber \\
\tau_\sigma \approx 1.27 \times 10^{23} \left( \dfrac{f}{10^{10}\text{~GeV}} \right)^5\text{~sec}.
\eea


\vskip .2in
\noindent
{\bf III. Technical naturalness of the ultra-weak sector}
\vskip .2in
\noindent

The ultra-weak couplings that have been introduced are technically natural.
In general, suppose we have a theory with various fields $\sigma_i$, $\phi_i$ with ``large'' couplings $\lambda_i \sim {\cal{O}}(1) $ and ultra-weak couplings $\zeta_i \ll {\cal{O}}(1)$.  The theory is defined by  a classical potential
\beq
V(\sigma, \phi_i, \lambda_i, \zeta_i)    =  V_1(\phi_i, \lambda_i) + V_2(\sigma_i, \phi_i, \zeta_i).
\eeq
Here the full potential decomposes into components
$V_1$ and $V_2$ where  $\frac{\delta}{\delta \sigma_i}V_1 = \frac{\delta}{\delta \zeta_i}V_1 = 0 $,
and $\frac{\delta}{\delta \lambda_i}V_2 = 0 $.

The RG equations for the $\zeta_i$ will then take the form
\beq
\frac{d\zeta_i}{d\ln(\mu)}= \beta_{\zeta_i} = \sum_{\zeta_j} \zeta_j F_i^j(\zeta_i, \lambda),
\eeq
with polynomial functions $ F_i^j(\zeta_i, \lambda_i) $.
The set of couplings $\{\zeta_i \}$ is {\em multiplicatively renormalized} and  the $\zeta_i$
can therefore be {\em technically naturally small}.  

This  multiplicative renormalization of the $\zeta_i$ arises because the fields $\sigma_i$ are
associated with approximate shift symmetries $\sigma_i \rightarrow \sigma + \epsilon_i f$ of the action (see Appendix). The smallness of the couplings $\zeta_i$ are protected by the shift, \ie, the 't~Hooft naturalness condition $\zeta_i \ll 1$ is satisfied since, in the limit $\zeta_i\rightarrow 0$, we have an enhanced
exact shift symmetry of the action.  Small $\zeta_i$ represents a small breaking
of this symmetry.

Given that the scale of gauge couplings in the SM is ${\cal{O}}(1)$, 
the shift symmetry limit can exist only if the $\sigma_i$ are gauge singlet fields. 
Indeed, it is not meaningful to talk about shift symmetries for 
fields that carry gauge charges such as the Higgs boson (unless one is interested in the consequences
of dynamics in the limit that gauge couplings can be ignored).  The couplings $\lambda_i$ of
fields such as the Higgs boson will receive additive corrections from gauge couplings and will not be
multiplicatively renormalized.  They will run according to the RG and become comparable  in size to the gauge couplings.

Of course, our argument is subject to gravitational effects.
All fields including $\sigma$ couple to gravity, which
is a gauge theory, so the condition of ultra-weak $\zeta_i$
couplings is subject to whether or not the shift symmetry can be maintained in the context of gravity.  This can be done if
the contributions to the RG equations
from  conformal couplings $\xi_i$, which appear in terms like $\half \xi \sigma ^2 R$, 
can remain ultra-weak.  These, in turn, will involve effective gravitational
couplings, an example of which is the recent ``Agravity'' model of Salvio and Strumia \cite{Salvio}.
It does appear possible to maintain the ultra-weak limit of the $\zeta_i$ within the
context of this scheme, and if the gravitational corrections are responsible for generating the $\zeta_{i}$ then a simple explanation for the hierarchy between $\zeta_2$ and $\zeta_1$ may be possible. If instanton effects are relevant and yield additive corrections to the $\zeta_i$,
we expect these to be suppressed as $\exp(-8\pi^2/\zeta_i)$. 

Hence, the shift symmetry may be a powerful constraint that admits a natural
sector of ultra-weakly coupled physics.  

\vskip .2in
\noindent
\z{\bf IV. Classical scale invariance}
\vskip .1in
\noindent
Up to now we have assumed that the theory obeys classical scale invariance in the sense that scale invariance is broken only through the trace anomaly. This assumes, as is the case in dimensional regularization, that the radiative corrections to scalar masses that are quadratically dependent on the cut-off scale are cancelled by the bare mass terms, leaving the scalars massless before spontaneous symmetry breaking. This makes sense in a pure field theory because only the renormalized masses are physical.  However, new physics at a high scale can spoil this by introducing contributions to the scalar masses that are proportional to the high scale. This is the case if there is a stage of Grand Unification, for which the contributions are proportional to the mass scale of the heavy GUT states, but can also happen even if there are no massive states, for example when the new scale is generated by the CW mechanism. 
In the model presented here, such corrections would affect the Higgs mass and give rise to the usual hierarchy problem, but they also affect the singlet state, despite its ultra-weak couplings, because a contribution to the $\sigma$ mass squared  of $O(\zeta_{i} \Lambda^2)$ will dominate over the CW potential for $\Lambda>O(\text{TeV})$. To avoid this we envisage two possibilities.

\z{The first is that there are no high scales of the type discussed above. Of course this cannot be true if gravity is included, but, as discussed above, it may be that gravity respects the shift symmetry and the gravitational corrections to the dilaton mass are small. However, one would still expect an unacceptably large contribution to the Higgs mass, thereby reintroducing the hierarchy problem.  Alternatively, if the model is UV complete so that it does not have Landau poles, gravity may not contribute to the scalar masses at all \cite{Dubovsky:2013ira}.  This case is analogous to that of a pure field theory with classical scale invariance and guarantees that the scalar sector remains massless in the absence of spontaneous symmetry breaking.}

\z{The second possibility is to super-symmetrize the model so that the quadratic mass terms have a low SUSY scale cut-off. In this case, one {\it can} have a stage of Grand Unification without introducing unacceptably large scalar mass contributions. A supersymmetric version of the model requires an additional Higgs doublet that somewhat complicates the model. We will discuss this possibility in detail in a partner paper that considers the mechanism in the context of axion solutions to the strong CP problem.}

\vskip .2in
\noindent
{\bf V. Conclusions}
\vskip .1in
\noindent

We have considered the possibility that the Higgs boson mass arises from an ultra-weak
sector that contains an effective dilaton.  The dilaton emerges with a very small mass
and couples (with rescaled couplings) to all final states accessible to the Higgs boson.

The ultra-weak sector is technically natural and is protected by a shift symmetry.
We believe this symmetry can be maintained in quantum gravity.

In a parallel work \cite{AHR}, we will incorporate the axion, which fits naturally into an ultra-weak complex singlet field generalization of this idea.  We will discuss further cosmological 
and phenomenological implications therein.

\vskip 0.5in
\noindent
{\bf Acknowledgements}
\vskip .2in
\z{For useful discussions, we thank
W. Bardeen and Giovanni Villadoro. Fermilab is operated by Fermi Research Alliance, LLC
under Contract No. DE-AC02-07CH11359 with the United
States Department of Energy. One of us (GGR) would like to thank the Leverhulme foundation for an emeritus fellowship without which this research would not have been initiated.}

\vskip .5in
\noindent
{\bf Appendix: Shift Current and  the Dilaton}
\vskip .1in
\noindent
\renewcommand{\theequation}{A.\arabic{equation}}   
\setcounter{equation}{0}  

The field $\sigma$ with ultra-weak couplings is formally analogous to
a dilaton, as occurs in a spontaneous breaking of scale symmetry.  Let us
examine this relationship.

Spontaneous scale symmetry breaking can
be viewed in two ways. The conventional
description is to start with a scale invariant
theory, containing a dilaton with a shift-invariant potential,
and matter fields.  The dilaton's shift
symmetry is broken by the coupling to matter, \eg, as in Yukawa couplings. 
The stress-tensor is traceless.   The dilaton can
then acquire a nonzero VEV, and the matter fields then acquire mass, but
the stress tensor remains traceless. Hence, we end up with a scale invariant
theory, massive matter,  and a massless dilaton as the Nambu-Goldstone boson.

Alternatively, we can start with massive matter fields, and we include a dilaton with a shift-invariant potential,
but with couplings to matter that again break the shift symmetry. Now we compute the stress tensor and find that
it is {\em not traceless}, \ie, the scale current is not conserved. However, we can
find a linear combination of the scale current and the dilaton shift current 
that is conserved; the theory has a hidden
symmetry after all. 

To see this latter mode,
consider  an interacting massless scalar field and a massive fermion,
\beq
\label{theory10}
S=\int d^{4}x\left( \bar{\psi}i\slash{\partial}\psi + \frac{1}{2}\partial _{\mu }\sigma \partial
^{\mu }\sigma - V(\sigma,\psi) +{\cal{L}}_I\right),
\eeq
where
\beq
\label{V2}
V(\sigma, \psi)= m\bar{\psi}\psi + g\sigma\bar{\psi}\psi.
\eeq
${\cal{L}}_I = (-1/6) \partial ^{2}\sigma ^{2}$ 
is  an ``improvement term'' and does not affect the equations
of motion.
The usual diffeomorphism  $\delta x^\mu = \xi^\mu(x)$,
holding the metric fixed,  
then yields the ``improved stress tensor'' \cite{CCJ}
as $\delta S = (1/2)(\partial_{\{\mu} \xi_{\nu\}})T^{\mu\nu}$
(see Appendix A of \cite{CTH}):
\bea
\label{scalarfermion}
{T}_{\mu \nu } & = & \frac{2}{3} \partial
_{\mu }\sigma \partial_{\nu }\sigma -\frac{1}{6}\eta _{\mu \nu }\partial _{\rho
}\sigma \partial ^{\rho }\sigma -\frac{1}{3}\sigma \partial _{\mu }\partial
_{\upsilon }\sigma
\nonumber \\
& & 
\!\!\!\!\! \!\!\!\!\!\!\!\!\!\!\!\!\!\!
+\;\frac{1}{3}\eta _{\mu \nu }\sigma \partial ^{2}\sigma +
\frac{i}{2}\bar{\psi}\gamma_{\{\mu}\partial_{\nu\}}\psi+\eta _{\mu \nu }(V(\sigma) -i\bar{\psi}\slash\partial\psi ).
\nonumber \\ & &
\eea
The last term
can be dropped since it vanishes by the fermion equation of motion.

The scale current is derived by
  $\delta x^\mu = \epsilon(x) x^\mu$, yielding
$S_\mu = \delta S/\delta\partial^\mu\epsilon = x^\nu T_{\mu\nu}$. 
The divergence of $S^\mu$ is the trace of eq.(\ref{scalarfermion}),
\beq
\partial_\mu S^\mu={T}_{\mu}^{\mu } = m\bar{\psi}\psi,
\eeq
where equations of motion $ i\slash{\partial}\psi = m\psi +g\sigma\psi $   
and $\partial ^2 \sigma = -g\bar{\psi}\psi$ are used.
Therefore, we see that the scale symmetry is apparently broken by the fermion
mass.  

However,  there is a ``shift current'' for the field $\sigma$
defined by the ``shift transformation'' $\delta \sigma =  \epsilon f$,
where $f$ is some arbitrary mass scale. 
The shift transformation implies a Noether current  $ J_\mu^S = f\partial_\mu\sigma$.  
The $J_0^S \propto \dot{\sigma} $
component is the canonical momentum of $\sigma$, which induces operator shifts in the value of the 
field through the equal time commutation relations, much like a momentum operator $i\partial_\mu$ induces
shifts in position in ordinary quantum mechanics. The conservation law of $J^F_\mu$
is, of course, equivalent to the equation of motion of $\sigma$,  $\partial^\mu J_\mu^F = -f\frac{\delta}{\delta \sigma}V(\sigma)$,
where $V$ is a potential that depends  nonderivatively upon $\sigma$ and other
fields.

In the case of eq.(\ref{theory}), the shift
symmetry is broken by the Yukawa coupling since
\beq
\partial_\mu J_S^{\mu}= f\partial^2\sigma = -gf\bar{\psi}\psi.
\eeq
However, we see that with the special choice $gf = m$ we
have a conserved current $Q^\mu = S^\mu + J_S^\mu$, the sum of the shift current and the scale current
\beq
\partial_\mu\hat{Q}^\mu  = \left. (m-gf)\bar{\psi}\psi \rightarrow 0\right|_{gf = m}
\eeq
The theory therefore has a hidden symmetry. 

Note that we could obtain a
conserved scale current $\hat{S}^\mu$ by modifying the stress
tensor to
\beq
\hat{T}^{\mu\nu} = {T}^{\mu\nu}
-\frac{1}{3}\partial_{\{\mu }J^S_{\nu\}} +\;\frac{1}{3}\eta _{\mu \nu }\partial_{\rho}J_S^\rho.
\eeq
The modified stress tensor implies a modified scale current
$\hat{S}_\mu=  x^\nu \hat{T}_{\mu\nu}$ that has the trace
\beq
\partial_\mu \hat{S}^{\mu}=\hat{T}_{\mu}^{\mu} = \left. (m-gf)\bar{\psi}\psi \rightarrow 0\right|_{gf = m}.
\eeq

The modified stress tensor is precisely
what we would have obtained directly from the 
scale invariant theory,
\ie, eq.(\ref{theory10}, \ref{V2}) with $m=0$, 
and shifting $\sigma$ to a nonzero VEV
$\sigma \rightarrow \sigma + f$.    
The shift current is  playing a hidden role, buried in the stress tensor,
yielding the conserved scale current.



\begin{thebibliography}{99} 


\bibitem{bardeen}
  W.~A.~Bardeen,
  ``On naturalness in the standard model,''
  Fermilab-Conf-95-391-T, Fermilab-Conf-08-118-T  and private communication.\\
  C.~T.~Hill,
  ``Conjecture on the physical implications of the scale anomaly,''
  hep-th/0510177, presented at M. Gell-Mann 75th birthday celebration, Santa Fe, (2005).\\
  H.~Aoki and S.~Iso,
  Phys.\ Rev.\ D {\bf 86}, 013001 (2012)
  [arXiv:1201.0857 [hep-ph]].





\bibitem{general}



V. Elias, R. B. Mann, D. G. C. McKeon and T. G. Steele, 
Nucl. Phys. B 678 (2004) 147 [Erratum-ibid. B 703 (2004) 413];
  R.~Barbieri, L.~J.~Hall and V.~S.~Rychkov,
  Phys.\ Rev.\ D {\bf 74}, 015007 (2006);
  L.~Lopez Honorez, E.~Nezri, J.~F.~Oliver and M.~H.~G.~Tytgat,
  JCAP {\bf 0702}, 028 (2007);
  J.~R.~Espinosa and M.~Quiros,
  Phys.\ Rev.\ D {\bf 76}, 076004 (2007);
F. A. Chishtie, D. G. C. McKeon and T. G. Steele, 
Phys. Rev. D 77 (2008) 065007;\\
  R.~Foot, A.~Kobakhidze, K.~L.~McDonald and R.~R.~Volkas,
  Phys.\ Rev.\ D {\bf 77}, 035006 (2008);
F. A. Chishtie, T. Hanif, J. Jia, R. B. Mann, D. G. C. McKeon, 
T. N. Sherry and T. G. Steele, Phys. Rev. D 83 (2011) 105009;
  T.~Hur, D.~-W.~Jung, P.~Ko and J.~Y.~Lee,
  Phys.\ Lett.\ B {\bf 696}, 262 (2011)
  [arXiv:0709.1218 [hep-ph]];
  T.~Hambye and M.~H.~G.~Tytgat,
  Phys.\ Lett.\ B {\bf 659}, 651 (2008);
  L.~Alexander-Nunneley and A.~Pilaftsis,
  JHEP {\bf 1009}, 021 (2010);
  A.~Farzinnia, H.-J.~He and J.~Ren,
  Phys.\ Lett.\ B {\bf 727}, 141 (2013);
  T.~Hur and P.~Ko,
  Phys.\ Rev.\ Lett.\  {\bf 106}, 141802 (2011)
  [arXiv:1103.2571 [hep-ph]].
  A.~Arhrib, R.~Benbrik and N.~Gaur,
  Phys.\ Rev.\ D {\bf 85}, 095021 (2012);
  J.~S.~Lee and A.~Pilaftsis,
  Phys.\ Rev.\ D {\bf 86}, 035004 (2012);
  K.~Ishiwata,
  Phys.\ Lett.\ B {\bf 710}, 134 (2012);


\bibitem{Iso} 
  S.~Iso and Y.~Orikasa,
  PTEP {\bf 2013}, 023B08 (2013)
  [arXiv:1210.2848 [hep-ph]];
  C.~D.~Carone and R.~Ramos,
  Phys.\ Rev.\ D {\bf 88}, 055020 (2013);
T.G. Steele and Zhi-Wei Wang, Phys. Rev. Lett 110 (2013) 151601,\\ 
 T.~G.~Steele, Z.~-W.~Wang, D.~Contreras and R.~B.~Mann,
arXiv:1310.1960 [hep-ph];\\
  C.~Englert, J.~Jaeckel, V.~V.~Khoze and M.~Spannowsky,
  JHEP {\bf 1304}, 060 (2013);
  V.~V.~Khoze and G.~Ro,
  JHEP {\bf 1310}, 075 (2013);
  V.~V.~Khoze,  JHEP {\bf 1311}, 215 (2013);
  M.~Holthausen, J.~Kubo, K.~S.~Lim and M.~Lindner, JHEP {\bf 1312}, 076 (2013);
  R.~Foot, A.~Kobakhidze, K.~L.~McDonald and R.~R.~Volkas,
  arXiv:1310.0223 [hep-ph];
E.~Gabrielli, M.~Heikinheimo, K.~Kannike, A.~Racioppi, M.~Raidal and C.~Spethmann,
  arXiv:1309.6632 [hep-ph];
  M.~Aoki, S.~Kanemura and H.~Yokoya,
  Phys.\ Lett.\ B {\bf 725}, 302 (2013).
  T.~Hambye and A.~Strumia,
  Phys.\ Rev.\ D {\bf 88}, 055022 (2013),
R.~Dermisek, T.~Jung,  H.~Kim,
  arXiv:1308.0891 [hep-ph].


\bibitem{CTH} 
  C.~T.~Hill,
  arXiv:1401.4185 [hep-ph] (to appear in Phys. Rev. D)


\bibitem{CW} 
  S.~R.~Coleman and E.~J.~Weinberg,
  Phys.\ Rev.\ D {\bf 7}, 1888 (1973).
  %


\bibitem{AHR}  K. Allison, C.T. Hill and G. G. Ross, in preparation.

\bibitem{DFS} 
  M.~Dine, W.~Fischler and M.~Srednicki,
  Phys.\ Lett.\ B {\bf 104}, 199 (1981);
	A. P. Zhitnisky, Sov. J. Nucl. Phys. 31
	(1980) 260.
	
\bibitem{SVVZ} 
  J.~R.~Ellis, M.~K.~Gaillard and D.~V.~Nanopoulos,
  Nucl.\ Phys.\ B {\bf 106}, 292 (1976).\\
  M.~A.~Shifman, A.~I.~Vainshtein, M.~B.~Voloshin and V.~I.~Zakharov,
  Sov.\ J.\ Nucl.\ Phys.\  {\bf 30}, 711 (1979)
  [Yad.\ Fiz.\  {\bf 30}, 1368 (1979)].

\bibitem{Spira} 
  M.~Spira, A.~Djouadi, D.~Graudenz and P.~M.~Zerwas,
  Nucl.\ Phys.\ B {\bf 453}, 17 (1995)
  [hep-ph/9504378].

\bibitem{Salvio} 
  A.~Salvio and A.~Strumia,
  arXiv:1403.4226 [hep-ph]; this paper arrived during the course of
our present work, and presents an implicit ultra-weak sector with $f\sim M_\text{Planck}$.
  
\bibitem{Dubovsky:2013ira}
  S.~Dubovsky, V.~Gorbenko and M.~Mirbabayi,
  JHEP {\bf 1309} (2013) 045
  [arXiv:1305.6939 [hep-th]].



\bibitem{CCJ} 
  C.~G.~Callan, Jr., S.~R.~Coleman and R.~Jackiw,
  Annals Phys.\  {\bf 59}, 42 (1970).




\end{thebibliography}
\end{document}